\documentclass[12pt]{iopart}

\usepackage{graphicx}
\usepackage{hyperref}

\usepackage[utf8]{inputenc}
\usepackage[T1]{fontenc}
\begin{document}

\title{CLASH: Contrastive learning through alignment shifting to extract stimulus information from EEG}
\author{Bernd Accou, Hugo Van hamme, Tom Francart}

\address{KU Leuven, Belgium}
\ead{bernd.accou@kuleuven.be}

\begin{abstract}
Stimulus-evoked EEG data has a notoriously low signal-to-noise ratio and high inter-subject variability. 
We propose a novel paradigm for the self-supervised extraction of stimulus-related brain response data: a model is trained to extract similar information between two time-aligned segments of EEG in response to the same stimulus. The extracted information can subsequently be used to obtain better results in downstream tasks that utilize the response to the stimulus.
We show the efficacy of our method for a downstream task of decoding the speech envelope from auditory EEG. Our method outperforms other state-of-the-art denoising techniques, improving reconstruction scores by 45\%. Additionally, we show that in contrast to the baseline denoising techniques, our method can be used with data of unseen subjects and stimuli without retraining, improving decoding performance by 19\% and 34\% over raw EEG for two holdout datasets. Finally, the last experiment reveals that the accuracies obtained in the CLASH paradigm are significantly correlated with the percentile of obtained reconstruction correlation on the null distribution. In general, we showed that the proposed paradigm is suitable to train deep learning models to extract stimulus information from EEG while being stimulus feature agnostic.
\end{abstract}

\vspace{2pc}
\submitto{\JNE}
%
%
%

\section{Introduction}
Recorded brain responses to natural stimuli pose significant challenges in analysis due to the relatively low signal-to-noise ratio (SNR) (due to noise from neural processes unrelated to the stimulus, environmental noise, etc.) and high inter-subject variability (due to differences in anatomy, shifts in sensor locations, etc.).
A solution is to present a stimulus multiple times and average the responses, but this does not leverage the availability of multiple sensors. Methods like denoising source separation (DSS) \cite{sarela_denoising_2005, de_cheveigne_denoising_2008} with extensions such as time delayed denoising source separation (TDSS) \cite{de_cheveigne_time-shift_2010} and specific implementations such as joint decorrelation (JD) \cite{cheveigne_joint_2014} have been proposed to combine the spatial distribution of sensors to construct spatial filters that can disentangle noise sources from target signal sources. These techniques require the definition of a bias function. This bias function skews data after whitening in a direction of interest, e.g., stimulus-evoked responses. In EEG/MEG, the bias function is commonly defined as the stimulus-evoked response after averaging over epochs \cite{de_cheveigne_denoising_2008, cheveigne_joint_2014, de_cheveigne_time-shift_2010}, thus requiring repetitions of the exact same stimulus to the same subject.

Instead of utilizing a bias function, Canonical Correlation Analysis (CCA) \cite{hotelling_relations_1992, de_cheveigne_decoding_2018, dmochowski_extracting_2018, cheveigne_auditory_2021} can define a set of spatial filters for stimulus-evoked brain response data for a single repetition of a stimulus. CCA (and variations such as correlated component analysis) has been used to compare brain responses intra- and inter-subject to study engagement \cite{dmochowski_correlated_2012}, stimulus preference \cite{dmochowski_audience_2014} and speech intelligibility \cite{iotzov_eeg_2019}. Multi-way canonical correlation analysis (MCCA) \cite{de_cheveigne_multiway_2019} extends CCA to work with recordings of multiple subjects listening to the same stimulus. The downsides of MCCA are that it has to be retrained when new subjects are included, and a careful selection of shared components is necessary to obtain optimal results.
CCA can also directly relate a brain response to the stimulus \cite{de_cheveigne_decoding_2018, dmochowski_extracting_2018}, creating stimulus-aware spatial filters. Other methods, such as generalized eigenvalue decomposition (GEVD) \cite{lesenfants_data-driven_2019, das_stimulus-aware_2020}, can combine dimensionality reduction with stimulus-aware spatial filtering. Like the methods discussed previously, this is an efficient and interpretable method, but it requires a new model for different subjects/stimuli and has limited modeling capability.

Recently, self-supervised pre-training of neural networks has become a popular way to leverage unlabeled data for better downstream task performance in machine vision \cite{oord_representation_2019} and natural language processing \cite{devlin_bert_2019}. Generally speaking, models are first trained on a (pre-text) task to extract good general representations from large amounts of unlabeled data. After the pre-training stage, the representation model can be finetuned or used by another model to obtain better results on the actual task.  An example of self-supervised training is contrastive learning \cite{oord_representation_2019,chen_simple_2020,yang_simper_2023}. In the contrastive learning paradigm, a model learns by contrasting positive pairs (usually constructed by data augmentation from a chosen single example) and negative pairs (the chosen example and other unrelated elements in a batch) to encourage the model to map similar instances closer together in the embedding space. In the work of Yang et al. \cite{yang_simper_2023}, an extension over the typically used contrastive loss  (infoNCE \cite{oord_representation_2019}) is proposed, leveraging known similarity between the labels of the negative and anchor sample.

Previous cited methods (DSS, MCCA and contrastive learning) can be seen as a pre-text task, i.e., they can extract information in a self-or unsupervised manner, and the denoised brain response data can be used for other analysis. Given that the aim of this paper is to extract stimulus information from EEG evoked by continuous speech, a suitable downstream task would be decoding the speech envelope from EEG. For auditory tasks involving continuous speech stimuli, the decoding approach has been frequently used \cite{crosse_multivariate_2016,vanthornhout_speech_2018,di_liberto_low-frequency_2015}. In this approach, the speech envelope is decoded from EEG evoked by continuous speech with a linear model. The linear model will linearly combine multiple timesteps (also known as the integration window) and all channels of EEG to form a spatiotemporal filter that minimizes the mean-squared error between the predicted and actual speech envelope \cite{crosse_multivariate_2016,crosse_linear_2021}.

In this work, we propose a contrastive learning paradigm to extract stimulus information from stimulus-evoked brain responses in a self-supervised paradigm: Contrastive Learning through Alignment SHifting (CLASH). In this paradigm, the model is provided with two brain response segments evoked by the same stimulus. One of the segments (e.g., segment B) is shifted in time by a randomly chosen amount out of $N$ possible amounts before it is provided to the model. A non-linear deep neural network independently extracts the stimulus information from both brain responses into a representation space. Next, a classifier predicts which of the $N$ possible shifts was applied to segment B based on the extracted representations.
Our method is evaluated on an auditory EEG dataset using a subject-independent linear decoder to demonstrate the benefit of using the enhanced brain response data for downstream tasks. While an example is shown for EEG responses to natural speech stimuli, the same paradigm is applicable to other stimulus-evoked brain response tasks such as visual, motor or BCI-related tasks.

\section{Methods}

In this section, we provided background and implementation details  for our newly introduced CLASH paradigm, the baseline methods  (see section \ref{sec:denoising}) and the validation task (see section \ref{sec:validation}) used in this paper. Model weights and code can be found at \url{github.com/exporl/shift_detection}.

\subsection{Baseline methods}
\label{sec:denoising}
The current state-of-the-art in EEG denoising consists of linear filtering methods, such as denoising source separation (DSS) \cite{de_cheveigne_denoising_2008} and multiway canonical correlation analysis (MCCA) \cite{de_cheveigne_multiway_2019}. Both methods transform incoming EEG data to a component space, in which components are ranked due to their relative power (DSS) or shared variance across subjects (MCCA). Denoised EEG is obtained by transforming components into component space, and truncating the lower-ranked components, followed by projecting back to sensor space.

In this paper, EEG denoised by MCCA and DSS were considered as baseline methods in section \ref{sec:experiments}. We used the implementation of NoiseTools \cite{de_cheveigne_alain_noisetools_nodate} for both MCCA and DSS.

\subsubsection{DSS}
\label{sec:dss}

Denoising source separation (DSS) or Joint decorrelation (JD) defines an optimal spatial filter. In short, data is whitened by a principal component analysis (PCA), followed by applying the bias function. The bias function will bias components in a direction that will suppress artifacts and maximize the SNR. When applied to EEG, this bias function is typically chosen as the average across trials. Following the application of the bias function, a second PCA is performed, which results in $K$ DSS components. These transformations can be summarized by one transformation matrix. By calculating the pseudo-inverse of this transformation matrix, truncating the low-rank components and multiplying the transformation matrix with the pseudo-inverse, a denoising matrix is obtained. The optimal $K$ can be found by an exhaustive grid search\cite{de_cheveigne_denoising_2008,cheveigne_joint_2014}.

In our setup, we chose the average across subjects for each stimulus, as there are no repeated trials in our dataset. After a gridsearch over $2^K$ ($K = 0...6$) possible components, the $K$ components with optimal performance on the validation task were used for the experiments (see section \ref{sec:experiments}).

\subsubsection{MCCA}
\label{sec:mcca}
MCCA combines data from multiple subjects that listened to the same stimulus to obtain a filter that maximizes the shared brain responses across subjects \cite{cheveigne_multiway_2018}. Data is processed independently for each recording by submitting it to a PCA, followed by concatenation in the channel dimension and a final PCA across all data. As with DSS, this operation can be summarized by one transformation matrix, transforming the whole dataset into Shared component (SC) space. Shared components represent the shared brain responses across all subjects. Moreover, the transformation matrix can be split up into subject-specific transformation matrices, transforming the data of 1 subject into canonical components (CC). The SCs are the sum of the CCs across subjects. Data can be denoised by transforming data to component space, calculating the pseudo-inverse, truncating low-rank components and multiplying the truncated transformation matrix with the truncated pseudo-inverse.

In our setup, we performed a gridsearch, which considered 40 or 64 components after the first PCA and (1, 4, 8, 16, 32, 40, 64, 128, 200, 500, 1000, 2000, all) components after the second PCA. After the gridsearch the optimal combination of PC's was chosen based on the reconstruction score on the validation task.

\subsubsection{Benefits and limitations}
Both DSS and MCCA have been shown to be successful denoising and analysis tools in multiple studies. Due to their linear nature, they are easily and relatively quickly trainable, as well as understandable for analysis of suppressed artifacts and stimulus-evoked potentials.

For denoising purposes, however, they carry some limitations. Because they are linear spatial filters, they can only linearly combine channels, not allowing spectro-temporal processing. More importantly, both MCCA and DSS create subject-specific filters, meaning a different filter has to be computed for new subjects, requiring recomputation when new subjects are acquired and the bulk storage of sensitive brain imaging data, limiting applicability outside research. We aim to solve these limitations in our proposed method (see section \ref{sec:paradigm}).

\subsection{CLASH}
\label{sec:paradigm}

The main goal of the CLASH paradigm is to extract stimulus-related information from EEG in a stimulus-feature agnostic and subject- and stimulus-independent manner. To train a model in this paradigm, 2 time-aligned EEG segments evoked by the same stimulus are provided to the input to of a stimulus information extraction model (see Figure \ref{fig:paradigm}, the inputs at the left of the figure). One of the EEG segments is shifted temporally by a randomly chosen shift $S$ from a set of $N$ possible shifts (see the left-most rectangle in Figure \ref{fig:paradigm}). The two segments are then provided to a deep neural network that independently extracts the stimulus-related information by transforming both segments into a representation space (the middle rectangle in Figure \ref{fig:paradigm}). The resulting representations of both segments retain the same time dimension as the input EEG and could be interpreted as a transformed multivariate time series. Next, both representations are submitted to a classifier that has to predict which shift $S$ out of $N$ possible shifts was applied before stimulus information extraction (the right-most rectangle in Figure \ref{fig:paradigm}). Assuming the only common activity in both EEG segments is the stimulus-related brain responses, the classifier can perform this task if the deep neural network creates representations with good temporal information of the stimulus. After training, the deep neural network can be used to extract stimulus information from EEG for other subjects/stimuli and for downstream tasks, such as speech decoding (see section \ref{sec:validation}). Note that the deep neural network does stimulus information extraction rather than denoising, i.e., the output of the deep neural network will contain time series that do not necessarily resemble the biological characteristics of the input EEG.

This paradigm is closely connected to contrastive learning approaches: the categorical classification is based on normalized similarity values (correlations), leading to a loss function (equation \ref{eq:loss}) similar to InfoNCE \cite{oord_representation_2019} or NT-XEnt \cite{chen_simple_2020}:
\begin{equation}
    Loss = -log(\frac{\exp(sim(z_s, z_ss)/\tau)}{\sum_{k=1,k\neq s}^N exp(sim(z_s, z_k)/\tau)})
    \label{eq:loss}
\end{equation}

Where $z_s$ in equation \ref{eq:loss} is the embedding of the previously shifted EEG segment, $z_ss$ is the embedding of the correctly shifted embedding of the non-shifted EEG segment, $z_k$ are the shifted embeddings of the non-shifted EEG segment, $sim$ is Pearson correlation and $\tau$ is the temperature of the softmax. While closely connected to contrastive learning approaches \cite{chen_simple_2020,oord_representation_2019}, a key difference here is that we create our positives ($z_ss$) and negatives ($z_k, k\neq ss$) in a strictly controlled manner from our anchor ($z_s$) instead of sampling random segments from the batch. This strict control allows us to also use a different loss, called SimPer loss \cite{yang_simper_2023}:

\begin{equation}
    SimPer Loss = - \sum_{j=1}^{N} \frac{exp(simlabel_{s,ss})}{\sum_{k=1,k\neq s}^{N} exp(simlabel_{s,k})} * Loss
    \label{eq:simper}
\end{equation}

With $simlabel$ in equation \ref{eq:simper} as a similarity metric between labels, $Loss$ as the loss defined in Equation \ref{eq:loss}. This effectively transforms the problem into a regression paradigm, as the relative similarity weights each positive-anchor pair. This loss has been shown to produce state-of-the-art performance in the video domain \cite{yang_simper_2023}.

\subsubsection{Implementation for this paper}
\label{sec:implementation}

\begin{figure}
    \centering
    \includegraphics[width=\textwidth]{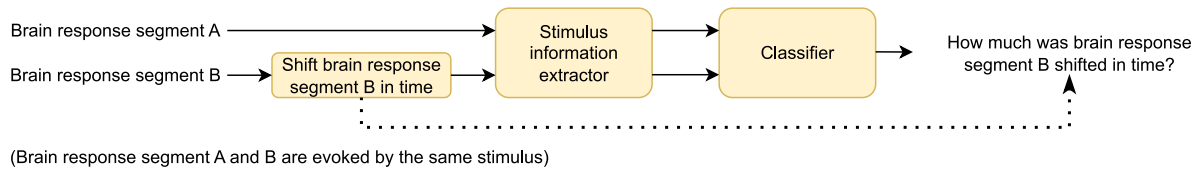}
    \caption{The general paradigm. In this paradigm, 2 segments of brain response data evoked by the same stimulus are presented as an input (segment A and B respectively). Segment B is then shifted in time by a randomly chosen amount of $N$ possible shifts. Next, the stimulus information extractor will independently non-linearly transform both segments to a representation space. Based on the representations, a classifier model will then classify how much segment B was shifted with regards to segment A.}
    \label{fig:paradigm}
\end{figure}

In the implementation for this paper, we chose $N=13$ with possible shift values evenly spaced (100~ms spacing) around zero (i.e. -600~ms, -500~ms, ..., 500~ms, 600~ms). This corresponds to the left-most yellow rectangle in Figure \ref{fig:paradigm}.

\begin{figure}
    \centering
    \includegraphics[width=\textwidth]{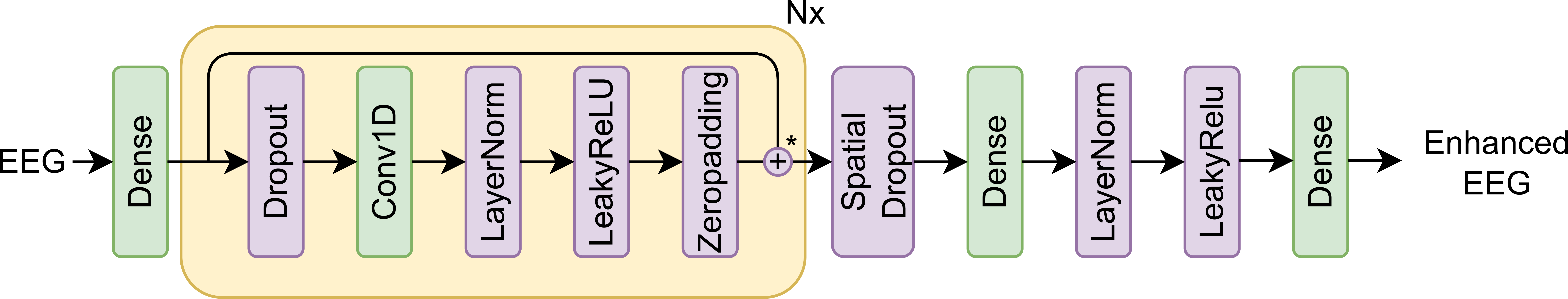}
    \caption{A CNN model, based on the VLAAI network \cite{accou_decoding_2023}. In our experiments, $N$=5.}
    \label{fig:shift_cnn}
\end{figure}

As the stimulus information extractor (the middle rectangle in Figure \ref{fig:paradigm}, we chose a convolutional neural network (CNN) based on the architecture of the VLAAI model \cite{accou_decoding_2023}, as displayed in Figure \ref{fig:shift_cnn}. This model consists of a dense layer, transforming the input 64 channels into 256 channels. Following the first dense layer, a block is used that is repeated $Nx$ = 5 times. This block consists of a dropout layer \cite{srivastava_dropout_2014} with a dropout rate of 10\%, followed by a Conv1D layer with 256 filters and a kernel size of 8. After the Conv1D, layer normalization \cite{ba_layer_2016}, a LeakyReLU \cite{maas_rectier_2013} non-linearity and zero-padding in the time dimension with 1 sample in from and 6 samples in the back is applied. Finally, a skip connection adds the input before the dropout layer to the output of the zero-padding layer, except in the last instance of this block, where the skip connection is omitted. Following the repeatable block, spatial dropout \cite{tompson_efficient_2015} with a dropout rate of 20\% is applied. Finally, a Dense layer with a dimension of 1024 is used, followed by layer normalization \cite{ba_layer_2016}, a LeakyRelU \cite{maas_rectier_2013} and a final Dense layer with a dimension of $D$. Note that the main scope of this paper is a proof of concept of the proposed paradigm, not necessarily developing the best possible architecture for the stimulus information extractor. Therefore, we consider exhaustive architecture searches out of the scope of this paper. The hyperparameters were chosen based on non-exhaustive searches and the original VLAAI architecture \cite{accou_decoding_2023}.

After the stimulus information extractor has transformed both EEG segments, the classifier (the right-most rectangle in Figure \ref{fig:paradigm}) predicts how much segment B was shifted with regards to segment A (see also Figure \ref{fig:paradigm}). Concretely, the classifier shifts the transformation of segment B by all $N$ = 13 possible shifts and correlates each of the newly obtained shifted segments with the transformed segment A for each channel separately. The correlations are averaged across the channel dimension, resulting in $N$=13 correlations that will be submitted to a softmax to obtain the model's final prediction. Note that this classifier implementation has no trainable parameters.

The model was trained on all possible pairwise combinations of recordings of subjects listening to the same stimulus, using 10-second segments without overlap in batches of 64 segments. An output dimension of $D=64$, the loss of equation \ref{eq:simper} with  $\frac{1}{\frac{x}{N}+1}$ as $simlabel$ function and a softmax temperature of 0.01 were used unless otherwise specified. The Adam \cite{kingma_adam_2015} optimizer was used with a learning rate of $10^{-3}$. Early stopping was applied with a patience factor of x. The model was implemented in TensorFlow version 2.10 \cite{abadi_tensorflow_2015}. Code and model weights are available at \url{github.com/exporl/shift_detection}.


\subsection{Validaton method}
\label{sec:validation}
The most common way to relate speech to continuous speech is by reconstructing the speech envelope from the EEG. This can be done using a linear decoder, which linearly combines all channels and a number of time-lagged versions from the EEG to obtain the reconstructed envelope. This model can be trained using ridge regression \cite{lalor_vespa_2006,crosse_multivariate_2016,crosse_linear_2021}.

To evaluate the methods proposed in this paper, we trained a subject-independent linear decoder with an integration window (corresponding to time lags of the EEG) from -100ms to 400ms. 3-fold cross- validation was performed across recordings to select the optimal ridge regression regularisation ($\lambda$) value (from a range of $10^x, x=[-6, -5, ..., 5, 6]$). After cross-validation, a model is trained across all folds using the optimal $\lambda$. Finally, the model is evaluated on the test set of each recording. For significance testing, the model's predictions were circularly shifted 100 times by a random amount with regard to the actual speech envelope for each recording to obtain a null distribution. The percentile of the mean correlation for each model on the null distribution was taken as the p-value. This model was trained across all stimuli for the stimulus information extractor CNN, DSS and MCCA.

\subsection{Dataset}
We re-used the data from Accou et al. \cite{accou_decoding_2023}. In brief, the dataset contains data from 80 native Flemish speakers (18-30 years old) who listened to 2-8 (on average 6) stories narrated in Dutch while their EEG was recorded. A holdout set of 26 subjects is also available. Each story is approximately 15 minutes in duration. All participants gave informed consent for this study, approved by the local Medical Ethics Committee. This dataset contains approximately 188 hours of EEG recordings in total, on average 1 hour and 46 minutes per subject.

To evaluate the generalization of the stimulus information extractor and CLASH paradigm more thoroughly, the single-speaker data from the dataset of Fuglsang et al. \cite{fuglsang_noise-robust_2017} was used. This dataset contains 10 trials of 50 seconds of EEG data elicited by narrated audiobooks in Danish by a single speaker per subject (18 subjects in total). Like the dataset used in Accou et al. \cite{accou_decoding_2023}, 64 channel EEG is recorded with a BioSemi ActiveTwo system.

Each EEG recording and the corresponding stimulus was split into 80\% training, 10\% validation and 10\% test set. The validate and test set were extracted from the middle of the recording to avoid artifacts and attention deviations of the participant that may occur at the beginning and end of the story.

\subsection{Preprocessing}
EEG was highpass filtered at 0.5 Hz with a non-causal 4th-order Butterworth filter. Subsequently, EEG was downsampled to 1024, and a multi-channel Wiener filter \cite{somers_generic_2018} was applied to remove eyeblinks. Next, EEG was re-referenced to a common average and downsampled to 64 Hz. 
The speech envelope was extracted from the stimuli using a gammatone filterbank, followed by downsampling to 64Hz, in a similar procedure as  \cite{accou_decoding_2023}. The stimuli were only used for the validation task.
Both EEG and speech envelope were split into 3 sets per recording: train, validate and test, using respectively 80\%, 10\% and 10\% from the recording data. The validate and test set were extracted from the middle of the recording. Data was standardized by computing the mean and standard deviation across time for the train set, followed by subtracting the computed mean and dividing by the standard deviation on the train, validation and test set. 

\section{Experiments}
\label{sec:experiments}
In this section, we demonstrated the efficacy of CLASH and the trained stimulus information extractor. In our experiments, we focused on the speech envelope decoding tasks, as this is the primary target for the training dataset. In the first experiment, we compared our method to DSS and MCCA. Next, we investigated generalization to unseen subjects and stimuli. Finally, we evaluated the influence of different loss functions and compared the CLASH accuracy to the obtained correlations in the validation task.

\subsection{Comparison to baselines}
\label{sec:comparison}
DSS, MCCA and the stimulus information extractor trained in the CLASH paradigm were trained and evaluated on the train set. After training, all methods were applied to the data before training and evaluating a subject-independent linear decoder. As explained in section \ref{sec:dss} and \ref{sec:mcca}, an optimal number of components was found for DSS and MCCA through a gridsearch. For DSS, 4 components were retained. For MCCA, 40 components were retained after the first PCA and 4 components after the second PCA. 

\begin{figure}
    \centering
    \includegraphics[width=\textwidth]{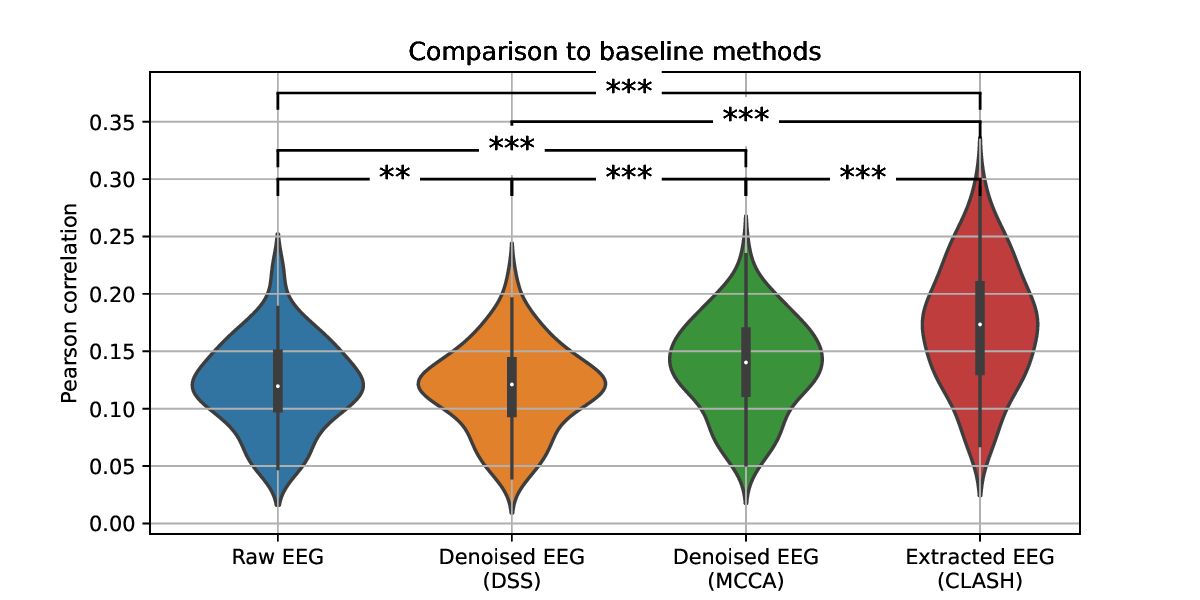}
    \caption{Comparison between a linear decoder trained and evaluated on raw/denoised/enhanced EEG as an input. Each point in the violin plot is the reconstruction score for one subject on the test set, averaged across recordings. Our proposed model significantly outperforms all baseline methods (p $< 10^-9$).}
    \label{fig:baselines}
\end{figure}

As shown in Figure \ref{fig:baselines} (a), our proposed method significantly outperforms every method except the shared components of MCCA. Compared to using raw EEG, our method improves significantly (from 0.120 to 0.173 median Pearson correlation, a relative increase of 45\% in median performance, Wilcoxon signed rank test with Holm-Bonferroni correction, p $<$ 0.001). For 79 of the 80 subjects, an improvement in the reconstruction score was found. For all methods, significant neural tracking was found after comparison to the null distribution (p=$1*10^{-3}$, p=$2*10^{-3}$ p=$7*10^{-4}$ and p=$2*10^{-4}$ for raw EEG, denoised EEG by DSS and MCCA, and stimulus information extracted EEG with CLASH respectively, see also section~\ref{sec:validation}).

\subsection{Generalization}

\begin{figure}
    \centering
    \includegraphics[width=\textwidth]{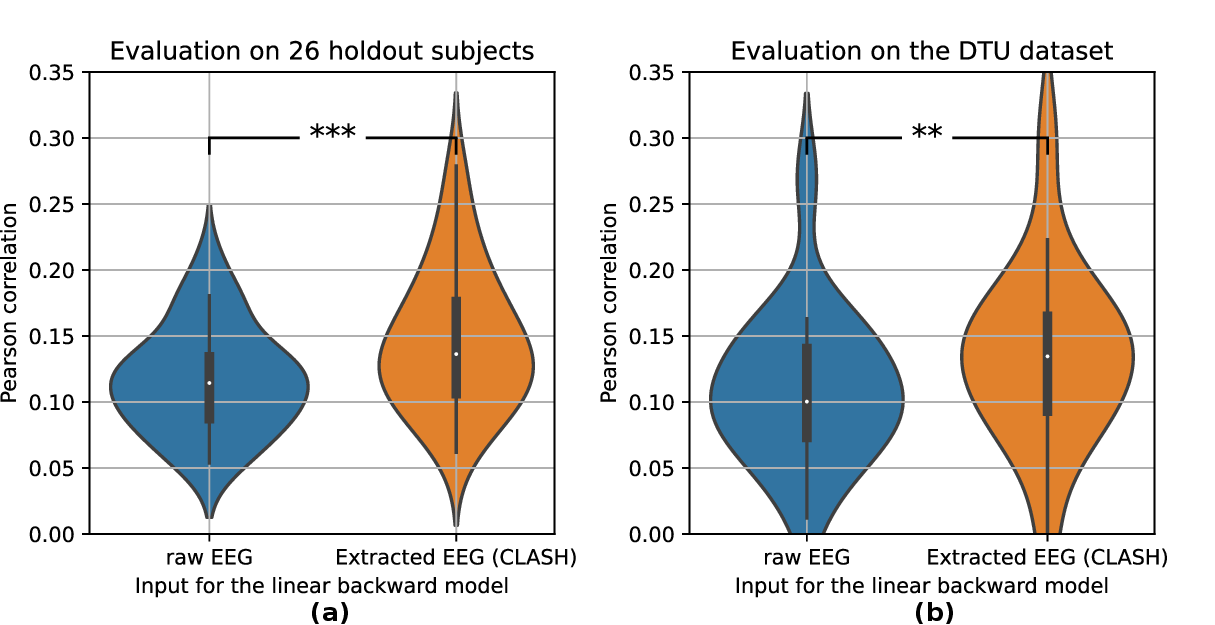}
    \caption{The stimulus information extractor and the corresponding linear decoder of Figure \ref{fig:baselines} are evaluated on 2 holdout datasets and compared to the linear decoder with raw EEG of Figure \ref{fig:baselines}. The holdout dataset in \textbf{(a)} contains 26 subjects not included in the train set, who listened to the same stimuli as the normal test set. The DTU dataset in \textbf{(b)} contains 18 unseen subjects that listened to unseen audiobooks. CLASH significantly outperforms the linear decoder in both scenarios. Note that DSS/MCCA has to be retrained when new subjects and/or stimuli are used and are therefore not included in this analysis. Each point in the violin plot is the reconstruction score for one subject on the test set, averaged across recordings. 
    \\(n.s.: p $\geq$ 0.05, *: 0.01 $\leq$ p < 0.05, **: 0.001 $\leq$ p < 0.01, ***: p < 0.001)
    }
    \label{fig:generalization}
\end{figure}

In contrast to DSS and MCCA baseline methods, our model could be used for unseen subjects and stimuli. In this experiment, we compared our model to raw EEG for the holdout dataset containing 26 unseen subjects but seen stimuli and  the DTU dataset containing both unseen subjects and stimuli. The already computed subject-independent decoders from section \ref{sec:comparison} for raw EEG and the stimulus information extractor were used. The results are displayed in Figure \ref{fig:generalization}.  The representations generated by the stimulus information extractor significantly outperform the raw EEG for both holdout datasets (p < 0.01). Compared to the normal test set, however, the obtained scores are significantly lower (reduction of 21\% and 22\% for the holdout and DTU dataset, respectively. Wilcoxon rank-sum test with Holm-Bonferroni correction, p=0.04 for both). This lower score could be attributed to slight overfitting on the training subjects and/or stimuli or to the inherent differences in difficulty in the detection of neural tracking for the different datasets. The percentile of the mean correlation score for the raw EEG method has decreased from 99.9 to 99.8 and 98.0 (i.e., p=0.002 and p=0.02) for the test set compared to the holdout and DTU dataset, respectively, indicating that it is harder to detect neural tracking for the DTU dataset. Significant neural tracking was also found for the stimulus information extractor method (p=0.002 and p=0.03, respectively).

\subsection{Loss function}
\label{sec:loss}
As stated in section \ref{sec:paradigm}, the controlled approach of generating negatives allows for different loss functions, such as SimPer Loss\cite{yang_simper_2023}. Moreover, different $simlabel$ functions for SimPer can be defined.

\begin{figure}
    \centering
    \includegraphics[width=\textwidth]{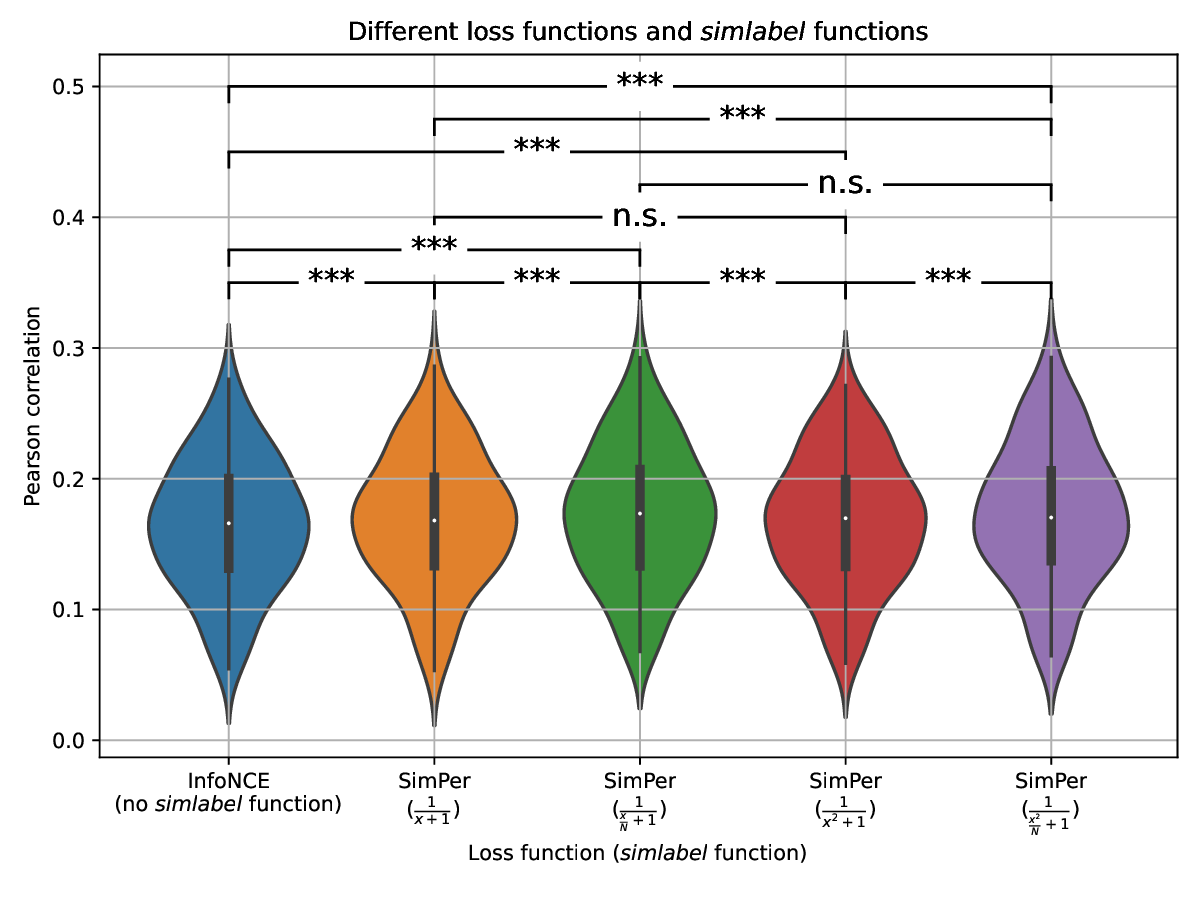}
    \caption{Different label similarity functions (see $simlabel$ in equation \ref{eq:simper}) were used when training the model with SimPer loss \cite{yang_simper_2023} . All the proposed distance functions improve significantly over the InfoNCE loss (p < 0.001). The optimal $simlabel$ function was $\frac{1}{\frac{x^2}{N}+1}$ with a median Pearson correlation of 0.167.\\(n.s.: p $\geq$ 0.05, *: 0.01 $\leq$ p < 0.05, **: 0.001 $\leq$ p < 0.01, ***: p < 0.001)}

    \label{fig:loss}
\end{figure}

In this experiment, the influence of 4 $simlabel$ functions on SimPer loss (see equation \ref{eq:simper}) were explored:
$\frac{1}{x+1}$, $\frac{1}{\frac{x}{N}+1}$, $\frac{1}{x^2 + 1}$, $\frac{1}{\frac{x^2}{N}+1}$. The results are shown in Figure \ref{fig:loss}.  A Wilcoxon signed-rank test with Holm-Bonferroni correction was used to determine if the differences between results were significant. All the proposed distance functions improve significantly over the infoNCE loss (p < 0.001). The optimal distance functions were  $\frac{1}{\frac{x}{N}+1}$ and $\frac{1}{\frac{x^2}{N}+1}$ with a median Pearson correlation of 0.173 and 0.170 respectively.

\subsection{Comparison of CLASH accuracy and reconstruction score}
A downside that is shared between the baseline methods and the CLASH paradigm is that the optimal hyperparameters have to be chosen based on the validation task, making hyperparameter optimization a complicated 2-step process. In this experiment, we evaluated whether there was a link between the accuracy of CLASH in detecting the correct shift vs. the reconstruction score (Pearson correlation) on the validation task. Therefore, we trained different hyperparameter tunings of the stimulus information extractor model and compared their obtained accuracy in CLASH with the obtained reconstruction score of a linear backward model trained on the extracted EEG. For possible hyperparameters, we started from the hyperparameters as specified in section \ref{sec:implementation} and varied one hyperparameter at a time. The temperature of the softmax was varied from 0.01 to [0.00001, 0.0001, 0.001, 0.1, 1.0]. The output channel dimension of the stimulus information extractor was varied from 64 to  [2, 4, 8, 16, 32, 128]. The loss functions were varied exactly as in section \ref{sec:loss}. In total, 16 different version of the stimulus information extractor were used.

\begin{figure}
    \centering
    \includegraphics[width=\textwidth]{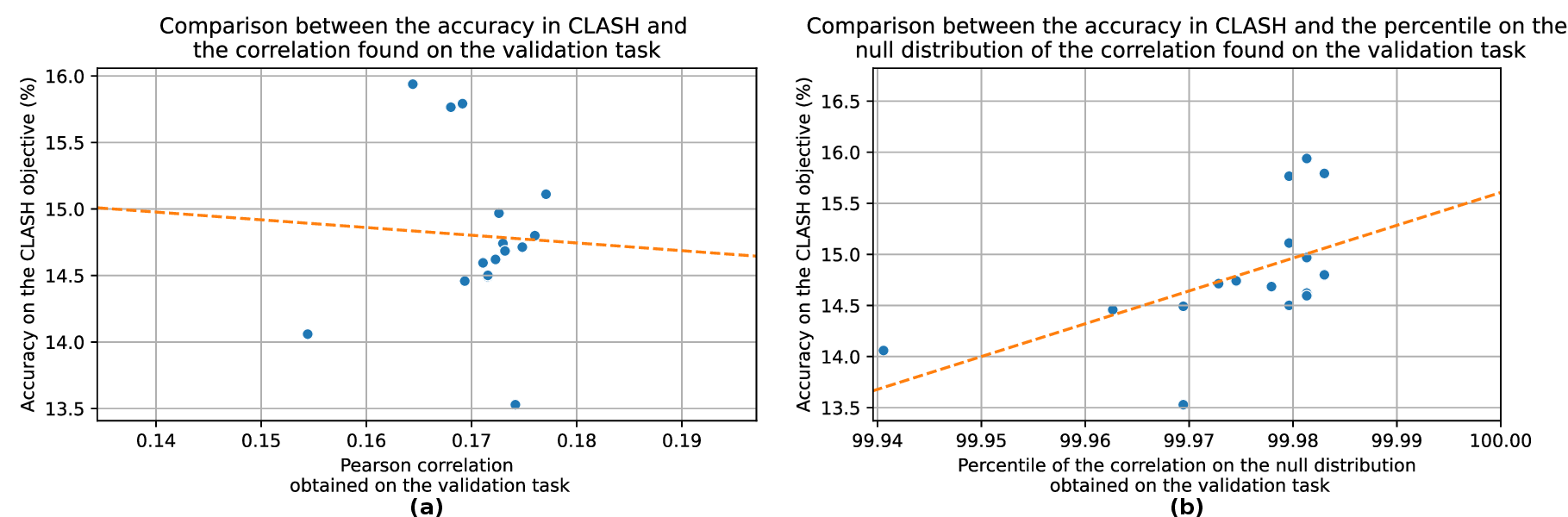}
    \caption{
    \textbf{(a):} The accuracy of CLASH is compared for different hyperparameter tunings of the extractor model to Pearson correlation score obtained by a linear backward model trained and evaluated on the representations of each extractor model. No significant correlation was found (Spearman correlation, $\rho$=0.0, p=1.0).\\
    \textbf{(b):} The accuracy of CLASH is compared for different hyperparameter tunings of the extractor model to the percentile of the Pearson correlation score on the null distribution obtained by a linear backward model trained and evaluated on the representations of each extractor model. In this case, a significant correlation was found (Spearman correlation, $\rho$=0.69, p=0.003), suggesting that the CLASH objective reflects the distance between the null and prediction distribution rather than the raw correlation values. }
    
    \label{fig:clash_vs_corr}
\end{figure}

In figure \ref{fig:clash_vs_corr} (a), the accuracy of the extractor models in the CLASH paradigm is compared to the obtained correlation scores on the validation task. No significant correlation was found between the CLASH accuracy and the reconstruction correlation (Spearman correlation, $\rho$=0.0, p=1.0). However, as shown in Figure \ref{fig:clash_vs_corr}
 (b), when comparing the CLASH accuracy with the percentile of the obtained correlation scores on the null distribution of the validation task, a significant negative correlation was found (Spearman correlation, $\rho$=0.69, p=0.003). This result suggests that the CLASH objective reflects the distance between the actual and null distribution rather than actual correlation values.
\section{Discussion}
The experiments showed that the CNN trained with CLASH significantly outperformed the baseline methods on the validation task by 45\%, 43\% and 23\% for raw EEG, EEG denoised by DSS and EEG denoised by MCCA, respectively (p < $10^{-9}$). In contrast to the baselines, our method can generalize to unseen subjects without any retraining, significantly outperforming the linear backward model using raw EEG by 19\% and 34\% on the holdout and DTU datasets, respectively.

The performance of DSS \cite{de_cheveigne_denoising_2008} seems rather low compared to raw EEG, possibly due to poor choice of bias function. The recommended bias function for DSS is the average of several presentations of the same stimulus within a subject. However, such data was not available for the chosen dataset. While MCCA \cite{de_cheveigne_decoding_2018} performed significantly better than raw EEG and DSS, it has to be retrained when new subjects are added. This implies the long-term storage of a large amount of data, substantial processing time per new subject, and, critically, the requirement of training data from each subject in response to the same stimulus as in the rest of the training set. This is not always feasible or even possible, particularly for practical applications outside of the research context. 

However, MCCA and DSS still have advantages over the proposed method. Firstly, when denoising data for interpretation, models trained with CLASH act like a black box due to the many non-linear transformations applied, in contrast to MCCA and DSS, which have been shown to produce interpretable filters. Moreover, the output of our model is not constrained to resemble EEG, so it is hard to make any biological interpretation of the signal, whereas this might be possible with MCCA and/or DSS. Secondly, our model is computationally much more expensive to train. In the layout described in section \ref{sec:implementation}, our CNN stimulus information extractor has approximately 1 million parameters, while DSS has $\#channels^{2}$ (without truncating) MCCA has $\#subjects * \#channels$ (without truncating). 

Our method showed fair generalizability to unseen subjects and stimuli, significantly improving decoding accuracy in the validation task over raw EEG. However, our model performed significantly worse on both the holdout and DTU datasets than the normal test set.  While we showed that it is harder to detect neural tracking for the DTU dataset, this does not seem to be the case for the holdout dataset, suggesting slight overfitting instead. This issue could potentially be alleviated by using a different and/or smaller architecture for the stimulus information extractor model and/or including more data for training. In general, contrastive learning methods require large amounts of data to obtain state-of-the-art performance (e.g., \cite{baevski_wav2vec_2020}). In our case, more data might be necessary, especially due to the low SNR of auditory EEG recordings.


In section \ref{sec:loss}, the SimPer loss \cite{yang_simper_2023} was shown to work well with all $simlabel$ candidates, showcasing its usefulness as a drop-in replacement for infoNCE when the label distance can be estimated. In our case, scaling the distance functions by the number of possible shifts delivered optimal performance.

In the last experiment, we show that the accuracies obtained in the CLASH paradigm are not significantly correlated with the obtained correlation scores on the validation task. However, a significant correlation was found between the CLASH accuracies and the percentile of the obtained correlation scores on the respective null distribution on the validation task. While for some applications, higher correlations might be desired to be able to e.g., clearly separate between groups and/or conditions or to reconstruct intelligible speech \cite{akbari_towards_2019}, the amount of neural tracking that a model can effectively extract should take an appropriate null distribution into account. Therefore, for modeling a robust relationship between EEG and the stimulus, the distance and overlap between the actual and null distribution should be maximized.

In general, we showed that the shift detection paradigm could effectively be used to train deep learning models to extract stimulus-related information from the EEG for downstream tasks. Further research is needed to validate whether this approach can be used in paradigms with other forms of complex stimulation (e.g., visual, watching a movie) or other modalities such as MEG.

\section{References}
\bibliography{references}

\providecommand{\newblock}{}
\begin{thebibliography}{10}
\expandafter\ifx\csname url\endcsname\relax
  \def\url#1{{\tt #1}}\fi
\expandafter\ifx\csname urlprefix\endcsname\relax\def\urlprefix{URL }\fi
\providecommand{\eprint}[2][]{\url{#2}}

\bibitem{sarela_denoising_2005}
Särelä J and Valpola H 2005 {\em Journal of Machine Learning Research\/} {\bf
  6} 233--272 ISSN 1533-7928
  \urlprefix\url{http://jmlr.org/papers/v6/sarela05a.html}

\bibitem{de_cheveigne_denoising_2008}
de~Cheveigné A and Simon J~Z 2008 {\em Journal of Neuroscience Methods\/} {\bf
  171} 331--339 ISSN 0165-0270
  \urlprefix\url{https://www.sciencedirect.com/science/article/pii/S0165027008002008}

\bibitem{de_cheveigne_time-shift_2010}
de~Cheveigné A 2010 {\em Journal of Neuroscience Methods\/} {\bf 189} 113--120
  ISSN 0165-0270
  \urlprefix\url{https://www.sciencedirect.com/science/article/pii/S0165027010001202}

\bibitem{cheveigne_joint_2014}
Cheveigné A~d and Parra L~C 2014 {\em NeuroImage\/} {\bf 98} 487--505 ISSN
  1053-8119 publisher: Academic Press
  \urlprefix\url{https://www.sciencedirect.com/science/article/pii/S1053811914004534}

\bibitem{hotelling_relations_1992}
Hotelling H 1992 Relations {Between} {Two} {Sets} of {Variates} {\em
  Breakthroughs in {Statistics}: {Methodology} and {Distribution}\/} Springer
  {Series} in {Statistics} ed Kotz S and Johnson N~L (New York, NY: Springer)
  pp 162--190 ISBN 978-1-4612-4380-9
  \urlprefix\url{https://doi.org/10.1007/978-1-4612-4380-9_14}

\bibitem{de_cheveigne_decoding_2018}
de~Cheveigné A, Wong D~D~E, Di~Liberto G~M, Hjortkjær J, Slaney M and Lalor E
  2018 {\em NeuroImage\/} {\bf 172} 206--216 ISSN 1053-8119
  \urlprefix\url{https://www.sciencedirect.com/science/article/pii/S1053811918300338}

\bibitem{dmochowski_extracting_2018}
Dmochowski J~P, Ki J~J, DeGuzman P, Sajda P and Parra L~C 2018 {\em
  NeuroImage\/} {\bf 180} 134--146 ISSN 1053-8119
  \urlprefix\url{https://www.sciencedirect.com/science/article/pii/S1053811917304299}

\bibitem{cheveigne_auditory_2021}
Cheveigné A~d, Slaney M, Fuglsang S~A and Hjortkjaer J 2021 {\em Journal of
  Neural Engineering\/} {\bf 18} 046040 ISSN 1741-2552 publisher: IOP
  Publishing \urlprefix\url{https://dx.doi.org/10.1088/1741-2552/abf771}

\bibitem{dmochowski_correlated_2012}
Dmochowski J, Sajda P, Dias J and Parra L 2012 {\em Frontiers in Human
  Neuroscience\/} {\bf 6} ISSN 1662-5161
  \urlprefix\url{https://www.frontiersin.org/articles/10.3389/fnhum.2012.00112}

\bibitem{dmochowski_audience_2014}
Dmochowski J~P, Bezdek M~A, Abelson B~P, Johnson J~S, Schumacher E~H and Parra
  L~C 2014 {\em Nature Communications\/} {\bf 5} 4567 ISSN 2041-1723 number: 1
  Publisher: Nature Publishing Group
  \urlprefix\url{https://www.nature.com/articles/ncomms5567}

\bibitem{iotzov_eeg_2019}
Iotzov I and Parra L~C 2019 {\em Journal of Neural Engineering\/} {\bf 16}
  036008 ISSN 1741-2552 publisher: IOP Publishing
  \urlprefix\url{https://doi.org/10.1088/1741-2552/ab07fe}

\bibitem{de_cheveigne_multiway_2019}
de~Cheveigné A, Di~Liberto G~M, Arzounian D, Wong D~D~E, Hjortkjær J,
  Fuglsang S and Parra L~C 2019 {\em NeuroImage\/} {\bf 186} 728--740 ISSN
  1053-8119
  \urlprefix\url{https://www.sciencedirect.com/science/article/pii/S1053811918321049}

\bibitem{lesenfants_data-driven_2019}
Lesenfants D, Vanthornhout J, Verschueren E and Francart T 2019 {\em Journal of
  Neural Engineering\/} {\bf 16} 066017 ISSN 1741-2552 publisher: IOP
  Publishing \urlprefix\url{https://doi.org/10.1088/1741-2552/ab3c92}

\bibitem{das_stimulus-aware_2020}
Das N, Vanthornhout J, Francart T and Bertrand A 2020 {\em NeuroImage\/} {\bf
  204} 116211 ISSN 1053-8119
  \urlprefix\url{https://www.sciencedirect.com/science/article/pii/S105381191930802X}

\bibitem{oord_representation_2019}
Oord A~v~d, Li Y and Vinyals O 2019 Representation {Learning} with
  {Contrastive} {Predictive} {Coding} arXiv:1807.03748 [cs, stat]
  \urlprefix\url{http://arxiv.org/abs/1807.03748}

\bibitem{devlin_bert_2019}
Devlin J, Chang M~W, Lee K and Toutanova K 2019 {BERT}: {Pre}-training of
  {Deep} {Bidirectional} {Transformers} for {Language} {Understanding}
  (Minneapolis, MN, USA: arXiv) arXiv:1810.04805 [cs]
  \urlprefix\url{http://arxiv.org/abs/1810.04805}

\bibitem{chen_simple_2020}
Chen T, Kornblith S, Norouzi M and Hinton G 2020 A {Simple} {Framework} for
  {Contrastive} {Learning} of {Visual} {Representations} arXiv:2002.05709 [cs,
  stat] \urlprefix\url{http://arxiv.org/abs/2002.05709}

\bibitem{yang_simper_2023}
Yang Y, Liu X, Wu J, Borac S, Katabi D, Poh M~Z and McDuff D 2023 {SimPer}:
  {Simple} {Self}-{Supervised} {Learning} of {Periodic} {Targets}
  arXiv:2210.03115 [cs] \urlprefix\url{http://arxiv.org/abs/2210.03115}

\bibitem{crosse_multivariate_2016}
Crosse M~J, Di~Liberto G~M, Bednar A and Lalor E~C 2016 {\em Frontiers in Human
  Neuroscience\/} {\bf 10} ISSN 1662-5161
  \urlprefix\url{https://www.frontiersin.org/articles/10.3389/fnhum.2016.00604}

\bibitem{vanthornhout_speech_2018}
Vanthornhout J, Decruy L, Wouters J, Simon J~Z and Francart T 2018 {\em Journal
  of the Association for Research in Otolaryngology\/} {\bf 19} 181--191 ISSN
  1438-7573 \urlprefix\url{https://doi.org/10.1007/s10162-018-0654-z}

\bibitem{di_liberto_low-frequency_2015}
Di~Liberto G~M, O’Sullivan J~A and Lalor E~C 2015 {\em Current Biology\/}
  {\bf 25} 2457--2465 ISSN 0960-9822
  \urlprefix\url{https://www.sciencedirect.com/science/article/pii/S0960982215010015}

\bibitem{crosse_linear_2021}
Crosse M~J, Zuk N~J, Di~Liberto G~M, Nidiffer A~R, Molholm S and Lalor E~C 2021
  {\em Frontiers in Neuroscience\/} {\bf 15} ISSN 1662-453X
  \urlprefix\url{https://www.frontiersin.org/articles/10.3389/fnins.2021.705621}

\bibitem{de_cheveigne_alain_noisetools_nodate}
{de Cheveigné, Alain} {NoiseTools}
  \urlprefix\url{http://audition.ens.fr/adc/NoiseTools/}

\bibitem{cheveigne_multiway_2018}
Cheveigne A~d, Liberto G~M~d, Arzounian D, Wong D, Hjortkjaer J, Fuglsang S~A
  and Parra L~C 2018 {\em bioRxiv\/}  344960
  \urlprefix\url{https://www.biorxiv.org/content/early/2018/06/12/344960}

\bibitem{accou_decoding_2023}
Accou B, Vanthornhout J, Van~hamme H and Francart T 2023 {\em Scientific
  Reports\/} {\bf 13} 812 ISSN 2045-2322 number: 1 Publisher: Nature Publishing
  Group \urlprefix\url{https://www.nature.com/articles/s41598-022-27332-2}

\bibitem{srivastava_dropout_2014}
Srivastava N, Hinton G, Krizhevsky A, Sutskever I and Salakhutdinov R 2014 {\em
  Journal of Machine Learning Research\/} {\bf 15} 1929--1958 ISSN 1533-7928
  \urlprefix\url{http://jmlr.org/papers/v15/srivastava14a.html}

\bibitem{ba_layer_2016}
Ba J~L, Kiros J~R and Hinton G~E 2016 Layer {Normalization} arXiv:1607.06450
  [cs, stat] \urlprefix\url{http://arxiv.org/abs/1607.06450}

\bibitem{maas_rectier_2013}
Maas A~L, Hannun A~Y and Ng A~Y 2013 Rectiﬁer {Nonlinearities} {Improve}
  {Neural} {Network} {Acoustic} {Models} {\em Proc. {ICML}\/} vol~30 p~6

\bibitem{tompson_efficient_2015}
Tompson J, Goroshin R, Jain A, LeCun Y and Bregler C 2015 Efficient {Object}
  {Localization} {Using} {Convolutional} {Networks} arXiv:1411.4280 [cs]
  \urlprefix\url{http://arxiv.org/abs/1411.4280}

\bibitem{kingma_adam_2015}
Kingma D~P and Ba J 2015 Adam: {A} {Method} for {Stochastic} {Optimization}
  {\em {ICLR}\/} \urlprefix\url{https://openreview.net/forum?id=8gmWwjFyLj}

\bibitem{abadi_tensorflow_2015}
Abadi M, Agarwal A, Barham P, Brevdo E, Chen Z, Citro C, Corrado G~S, Davis A,
  Dean J, Devin M, Ghemawat S, Goodfellow I, Harp A, Irving G, Isard M, Jia Y,
  Jozefowicz R, Kaiser L, Kudlur M, Levenberg J, Mane D, Monga R, Moore S,
  Murray D, Olah C, Schuster M, Shlens J, Steiner B, Sutskever I, Talwar K,
  Tucker P, Vanhoucke V, Vasudevan V, Viegas F, Vinyals O, Warden P, Wattenberg
  M, Wicke M, Yu Y and Zheng X 2015 {TensorFlow}: {Large}-{Scale} {Machine}
  {Learning} on {Heterogeneous} {Distributed} {Systems}

\bibitem{lalor_vespa_2006}
Lalor E~C, Pearlmutter B~A, Reilly R~B, McDarby G and Foxe J~J 2006 {\em
  NeuroImage\/} {\bf 32} 1549--1561 ISSN 1053-8119
  \urlprefix\url{https://www.sciencedirect.com/science/article/pii/S1053811906006434}

\bibitem{fuglsang_noise-robust_2017}
Fuglsang S~A, Dau T and Hjortkjær J 2017 {\em NeuroImage\/} {\bf 156} 435--444
  ISSN 1053-8119
  \urlprefix\url{https://www.sciencedirect.com/science/article/pii/S105381191730318X}

\bibitem{somers_generic_2018}
Somers B, Francart T and Bertrand A 2018 {\em Journal of Neural Engineering\/}
  {\bf 15} 036007 ISSN 1741-2552 publisher: IOP Publishing
  \urlprefix\url{https://doi.org/10.1088/1741-2552/aaac92}

\bibitem{baevski_wav2vec_2020}
Baevski A, Zhou H, Mohamed A and Auli M 2020 wav2vec 2.0: {A} {Framework} for
  {Self}-{Supervised} {Learning} of {Speech} {Representations} arXiv:2006.11477
  [cs, eess] \urlprefix\url{http://arxiv.org/abs/2006.11477}

\bibitem{akbari_towards_2019}
Akbari H, Khalighinejad B, Herrero J~L, Mehta A~D and Mesgarani N 2019 {\em
  Scientific Reports\/} {\bf 9} 874 ISSN 2045-2322 number: 1 Publisher: Nature
  Publishing Group
  \urlprefix\url{https://www.nature.com/articles/s41598-018-37359-z}

\end{thebibliography}

\end{document}